\documentclass[aps,preprint]{revtex4}
\usepackage{amsmath}
\usepackage{mathrsfs}
\usepackage{bm}

\begin{document}

\title{Nonlinear interaction between three kinetic Alfv\'{e}n waves}
\author{P. K. Shukla, G. Brodin and L. Stenflo \\
Department of Physics, Ume\aa\ University, SE-901 87 Ume\aa , Sweden}

\begin{abstract}
Using Hall-MHD theory, we focus on the resonant interaction between three kinetic Alfv\'{e}n waves. We thus derive three coupled equations which govern this process. It turns out that these equations contain the same coupling coefficient, directly showing that they satisfy the Manley-Rowe relations. The coupling coefficient can be written in a comparatively very simple form, that has not been deduced before. The decay rate, when a pump kinetic Alfv\'{e}n wave decays into two similar Alfv\'{e}n waves, is therefore easily estimated for plasmas of astrophysical interest. 
\end{abstract}

\pacs{52.35.Mw, 52.35.Bj}

\maketitle


The Alfv\'{e}n wave \cite{Alfven} has applications in most subfields of
plasma physics. Its linear properties has thus been considered in numerous
works (e.g. \cite{Hasegawa-Uberoi,Chian,Cramer}). Large amplitude circularly
polarized waves propagating along the external magnetic field $B_{0}\widehat{%
\mathbf{z}}$ have also been studied during several decades (e.g. \cite
{Sagdeev-Galeev,Stenflo1976,Goldstein1978,Derby1978,Shukla1985,Brodin1988})
whereas nonlinear Alfv\'{e}n waves propagating in other directions have only
been considered comparatively recently (e.g. \cite
{Shukla1985b,Petviashvili1985,Shukla1986,Petviashvili1992,Shukla-Stenflo1995,Pokhotelov1996,Voitenko1998,Shukla-Stenflo1999,Shukla2000,Stasi2000,Shukla-Stenflo2004,Shukla2004b,Wu-Chao2004,Voitenko2004,Shukla2005,Shukla2005b}%
). Of special interest \ is the kinetic Alfv\'{e}n wave, which can exist
when the plasma beta is larger than the electron to ion mass ratio $%
m_{e}/m_{i}$. Its frequency $\omega $ is 
\begin{equation}
\omega \approx k_{z}c_{A}\left[ 1+\frac{k_{\perp
}^{2}c_{s}^{2}-k_{z}^{2}c_{A}^{2}}{\omega _{ci}^{2}}\right] ^{1/2}
\label{Kinetic-DR}
\end{equation}
where the wavevector is $\mathbf{k=k}_{\perp }+k_{z}\widehat{\mathbf{z}}$, $%
c_{A}$ is the Alfv\'{e}n velocity, $c_{s}$ is a suitably normalized (e.g. 
\cite{Shukla-Stenflo1995}) ion acoustic velocity, and $\omega _{ci}$ is the
ion gyrofrequency.

It has been shown that kinetic Alfv\'{e}n waves can appear in plasmas, for
example in the ionosphere and magnetosphere, in the form of vortices (e.g. 
\cite{Shukla1985b,Petviashvili1985,Shukla1986,Petviashvili1992}). The
theoretical results (e.g. \cite{Shukla1986}) have been confirmed in
space-borne observations by means of the Intercosmos-Bulgarian-1300 and
Aureol-3 satellites \cite{Chmyrev1988,Chmyrev1991}, and the Cluster
satellites \cite{Sundqvist2005} with in situ simultaneous multi-point
measurements \cite{Chmyrev1988,Sundqvist2005}, as well as by the all-sky
camera ionospheric data \cite{Chmyrev1991}. The theoretical predictions and
estimates turned out to be in good agreement with the subsequent
observations (e.g. \cite{Chmyrev1988, Chmyrev1991,Sundqvist2005}).

It is however necessary to continue our studies of nonlinear Alfv\'{e}n
waves (e.g. other kinds of solitary structures \cite{Wu-Chao2004}),
three-wave decay mechanisms (e.g. \cite
{Voitenko1998,Voitenko2004,Shukla2005b,Sridhar}) to have the necessary
prerequisites for forthcoming investigations of other plasmas, in particular
for the solar corona (e.g. \cite{voitenko2000,Voitenko2002}) . That is the
reason why, in the present paper, we are going to use the Hall-MHD equations
to deduce the coupling coefficients for the interaction of three kinetic
Alfv\'{e}n waves.

The Hall-MHD equations can be written as 
\begin{equation}
\frac{\partial \rho }{\partial t}+\nabla \cdot (\rho \mathbf{v})=0
\end{equation}
\begin{equation}
\rho \frac{d\mathbf{v}}{dt}=-c_{s}^{2}\nabla \rho +\frac{(\nabla \times 
\mathbf{B})\times \mathbf{B}}{\mu _{0}}
\end{equation}
and 
\begin{equation}
\frac{\partial \mathbf{B}}{\partial t}=\nabla \times (\mathbf{v}\times 
\mathbf{B}-\frac{m_{i}}{e}\frac{d\mathbf{v}}{dt})
\end{equation}
where $d/dt=\partial /\partial t+\mathbf{v}\cdot \nabla $ , $e$ is the ion
charge, whereas $\rho $, $\mathbf{v}$, and $\mathbf{B}$ are the density,
velocity and magnetic field, respectively.

Considering the resonant interaction between three waves which satisfy the
matching conditions 
\begin{equation}
\omega _{3}=\omega _{1}+\omega _{2}
\end{equation}
and 
\begin{equation}
\mathbf{k}_{3}=\mathbf{k}_{1}+\mathbf{k}_{2}
\end{equation}
we can, using (1)-(5), derive the equations (see Ref. \cite
{Brodin-Stenflo1990} for details) 
\begin{equation}
\left( \frac{\partial }{\partial t}+\mathbf{v}_{g1,2}\cdot \nabla \right)
\rho _{1,2}=-\frac{1}{\partial \tilde{D}_{1,2}/\partial \omega _{1,2}}C\rho
_{2,1}^{\ast }\rho _{3}  \label{Coupling1}
\end{equation}
and 
\begin{equation}
\left( \frac{\partial }{\partial t}+\mathbf{v}_{g3}\cdot \nabla \right) \rho
_{3}=\frac{1}{\partial \tilde{D}_{3}/\partial \omega _{3}}C\rho _{1}\rho _{2}
\label{Coupling2}
\end{equation}
where 
\begin{eqnarray}
C &=&\frac{\omega _{1}\omega _{2}\omega _{3}}{\rho _{0}k_{1\perp
}^{2}k_{2\perp }^{2}k_{3\perp }^{2}}\left[ \frac{\mathbf{K}_{3}\cdot \mathbf{%
K}_{2}^{\ast }}{\omega _{1}}k_{1\perp }^{2}+\frac{\mathbf{K}_{3}\cdot 
\mathbf{K}_{1}^{\ast }}{\omega _{2}}k_{2\perp }^{2}+\frac{\mathbf{K}%
_{1}^{\ast }\cdot \mathbf{K}_{2}^{\ast }}{\omega _{3}}k_{3\perp }^{2}\right.
-  \notag \\
&&\frac{k_{1\perp }^{2}k_{2\perp }^{2}k_{3\perp }^{2}}{\omega _{1}\omega
_{2}\omega _{3}}c_{s}^{2}+\frac{i\omega _{ci}}{\omega _{3}}(\frac{k_{2z}}{%
\omega _{2}}-\frac{k_{1z}}{\omega _{1}})\left( (\mathbf{K}_{3}+\frac{i\omega
_{3}\mathbf{k}_{3}\times \mathbf{K}_{3}}{\omega _{ci}k_{3z}})\cdot (\mathbf{K%
}_{1}^{\ast }-\right.  \notag \\
&&\left. \left. \frac{i\omega _{1}\mathbf{k}_{1}\times \mathbf{K}_{1}^{\ast }%
}{\omega _{ci}k_{1z}})\times (\mathbf{K}_{2}^{\ast }-\frac{i\omega _{2}%
\mathbf{k}_{2}\times \mathbf{K}_{2}^{\ast }}{\omega _{ci}k_{2z}})-\mathbf{K}%
_{3}\cdot (\mathbf{K}_{1}^{\ast }\times \mathbf{K}_{2}^{\ast })\right) %
\right]  \label{Coefficient1} \\
&&  \notag
\end{eqnarray}
\begin{eqnarray}
\tilde{D}_{j} &=&\Biggl[\omega _{j}^{4}-\omega
_{j}^{2}k_{j}^{2}(c_{A}^{2}+c_{s}^{2})+k_{jz}^{2}k_{j}^{2}c_{A}^{2}c_{s}^{2}
\notag \\
&&-\frac{\omega _{j}^{2}k_{jz}^{2}k_{j}^{2}(\omega
_{j}^{2}-k_{j}^{2}c_{s}^{2})c_{A}^{4}}{\omega _{ci}^{2}(\omega
_{j}^{2}-k_{jz}^{2}c_{A}^{2})}\Biggr]\frac{(\omega
_{j}^{2}-k_{j}^{2}c_{s}^{2})}{\omega _{j}^{2}k_{j\perp
}^{2}k_{j}^{2}c_{A}^{2}}  \label{Diespersionfull}
\end{eqnarray}
and 
\begin{equation}
\mathbf{K}_{j}=\mathbf{k}_{j\perp }\frac{(\omega
_{j}^{2}-k_{jz}^{2}c_{s}^{2})}{\omega _{j}^{2}}+\frac{i\hat{\mathbf{z}}%
\times \mathbf{k}_{j\perp }(\omega
_{j}^{2}-k_{j}^{2}c_{s}^{2})k_{jz}^{2}c_{A}^{2}}{\omega _{ci}\omega
_{j}(\omega _{j}^{2}-k_{jz}^{2}c_{A}^{2})}+\frac{k_{j\perp
}^{2}k_{jz}c_{s}^{2}}{\omega _{j}^{2}}\hat{\mathbf{z}},  \label{Full-K}
\end{equation}
$\mathbf{v}_{gj}$ is the group velocity of wave $j$, and $c_{A}=(B_{0}/\mu
_{0}\rho _{0})^{1/2}$ . The derivation of (\ref{Coupling1}) and (\ref
{Coupling2}) is straightforward \cite{Brodin-Stenflo1990}. Our result has
the advantage that the same coupling coefficient C appears in both (\ref
{Coupling1}) and (\ref{Coupling2}). This means that the growth rates for $%
\omega _{j}>0$ will always be positive, as $\omega _{j}\partial
D_{j}/\partial \omega _{j}>0$. We could alternatively have used, instead of $%
\rho _{j}$, the velocity magnitudes ($v_{j}=\rho _{j}\omega _{j}\left| 
\mathbf{K}_{j}\right| /k_{j\perp }^{2}\rho _{0}$), which are more convenient
for the case where the waves are almost transverse.. For the particular case
of kinetic Alfv\'{e}n waves, in the limit $\omega \ll \omega _{ci}$, $%
k_{z}\ll k_{\perp },$ (\ref{Full-K}) can be approximated by 
\begin{equation}
\mathbf{K}_{j}\approx -i\frac{\omega _{ci}}{\omega _{j}}\hat{\mathbf{z}}%
\times \mathbf{k}_{j}.
\end{equation}
Thus, considering three kinetic Alfv\'{e}n waves, the interaction equations
can be rewritten as 
\begin{equation}
\left( \frac{\partial }{\partial t}+\mathbf{v}_{g1,2}\cdot \nabla \right)
v_{1,2}=-\omega _{1,2}C_{AAA}v_{2,1}^{\ast }v_{3}
\end{equation}
and 
\begin{equation}
\left( \frac{\partial }{\partial t}+\mathbf{v}_{g3}\cdot \nabla \right)
v_{3}=\omega _{3}C_{AAA}v_{1}v_{2}
\end{equation}
where the coupling coefficient can be approximated by 
\begin{equation*}
C_{AAA}=\frac{1}{2\omega _{ci}}\left[ k_{3\perp }\frac{\mathbf{k}_{1\perp
}\cdot \mathbf{k}_{2\perp }}{k_{1\perp }k_{2\perp }}-k_{2\perp }\frac{%
\mathbf{k}_{1\perp }\cdot \mathbf{k}_{3\perp }}{k_{1\perp }k_{3\perp }}%
-k_{1\perp }\frac{\mathbf{k}_{3\perp }\cdot \mathbf{k}_{2\perp }}{k_{3\perp
}k_{2\perp }}\right]
\end{equation*}
where $v_{j}$ is the magnitude of the velocity of wave $j$.

The present paper considers the decay of a pump kinetic Alfv\'{e}n wave $%
(\omega _{3},\mathbf{k}_{3})$ into two other kinetic Alfv\'{e}n waves.
Within Hall-MHD theory, the growth rate of the latter waves then turned out
to be of the order of $(k_{\bot }v_{3}/\omega _{ci})\omega _{3}$. It should
however be stressed that there are limitations on the validity of this
estimate. For example, we have here only considered a medium beta plasma
with $m_{e}/m_{i}\ll c_{s}^{2}/c_{A}^{2}\ll 1$, and adopted the Hall-MHD
model. We note that our result for the coupling strength does not coincide
with that of Ref. \cite{Voitenko1998} or Ref. \cite{Shukla2005b}. However,
this is to be expected as we are considering the regime $k_{z}c_{A}/\omega
_{ci}\gg k_{\bot }c/\omega _{pe}$, where $\omega _{pe}$ is the electron
plasma frequency, whereas previous authors have considered other regimes. In
any case, the nonlinear excitation of kinetic Alfv\'{e}n waves can
accelerate oxygen ions in the solar corona at heights of a few solar radii 
\cite{Voitenko2004}, and also contribute to the cross-field energization in
the auroral zones of the Earth%
\'{}%
s magnetosphere.

\end{document}